\documentclass[
 amsfonts,amsmath,amssymb,
 aps,
 prl,
  preprintnumbers,
  twocolumn,
 a4paper 
 ]
{revtex4}

\usepackage[utf8]{inputenc}

\usepackage{booktabs}
\usepackage{graphicx}
\usepackage{dcolumn}
\usepackage{bm}
\usepackage{upgreek}
\usepackage{MnSymbol}

\usepackage{xcolor}
\usepackage[colorlinks=true,
            linkcolor={red!50!black},
            urlcolor={green!50!black},
            citecolor={blue!80!black}]{hyperref}

\newcommand{\overbar}[1]{\mkern 1.5mu\overline{\mkern-1.5mu#1\mkern-1.5mu}\mkern 1.5mu}

\begin{document}
\preprint{QMUL-PH-21-07}

\title{Macdonald Indices for Four-dimensional $\mathcal N=3$ Theories}

\author{Prarit Agarwal} \email{agarwalprarit@gmail.com}
\author{Enrico Andriolo} \email{e.andriolo@qmul.ac.uk}
\author{Gergely K\'antor}\email{g.kantor@qmul.ac.uk}
\author{Constantinos Papageorgakis} \email{c.papageorgakis@qmul.ac.uk}
\affiliation{CRST and School of Physics and Astronomy, Queen Mary University of London,\\ Mile End Road, London E1 4NS, UK}
\date{\today}

\begin{abstract} 
  We brute-force evaluate the vacuum character for $\mathcal N=2$ vertex operator algebras labelled by crystallographic complex reflection groups $G(k,1,1)=\mathbb Z_k$, $k=3,4,6$, and $G(3,1,2)$. For $\mathbb Z_{3,4}$ and $G(3,1,2)$ these vacuum characters have been conjectured to respectively reproduce the Macdonald limit of the superconformal index for rank one and rank two S-fold $\mathcal N=3$ theories in four dimensions. For the $\mathbb Z_3$ case, and in the limit where the Macdonald index reduces to the Schur index, we find agreement with predictions from the literature.
\end{abstract}

\maketitle

\section{Introduction}

In recent years there has been intense activity pertaining to the study of superconformal theories (SCFTs) that do not admit a Lagrangian description. Theories with $\mathcal N\ge 2$ superconformal symmetry are ideal for such explorations. Despite the lack of perturbative control, one can still extract nontrivial data by exploiting the large amount of symmetry, e.g. by employing the power of dualities \cite{Argyres:2007cn,Gaiotto:2009we}, implementing the bootstrap programme \cite{Rattazzi:2008pe,Beem:2014zpa}, or evaluating superconformal indices \cite{Kinney:2005ej}.

In this context, ``pure'' $\mathcal N=3$ SCFTs---$\mathcal N=3$ theories which do not automatically enhance to $\mathcal N=4$---were envisioned in \cite{Ferrara:1998zt,Aharony:2015oyb} and engineered in string theory through the S$_k$-fold constructions of \cite{Garcia-Etxebarria:2015wns,Aharony:2016kai}. These are isolated, holographic SCFTs ($a_{\mathrm{4D}} = c_{\mathrm{4D}}$) with an F-theory dual on an $\mathrm{AdS}_5  \times (\mathrm{S}^5 \times \mathrm{T}^2)/\mathbb Z_k$ background. The gravity description was used in \cite{Imamura:2016abe,Arai:2019xmp} to evaluate the superconformal index in the large-rank limit. Candidates for additional rank-one and rank-two $\mathcal N=3$ examples were presented in \cite{Argyres:2016xua,Argyres:2019ngz}, by constructing corresponding Coulomb-branch geometries via gaugings of $\mathcal N=4$ theories by a discrete subgroup of the R-symmetry and electromagnetic duality groups. The ``Coulomb'' limit of the superconformal index \cite{Gadde:2011uv} and the Higgs-branch Hilbert series for these models were evaluated in \cite{Bourton:2018jwb}; see also \cite{Evtikhiev:2020yix}.

As $\mathcal N=3$ theories are automatically $\mathcal N=2$, a concrete computational handle can be established through the description of a ``Schur'' Bogomol'nyi--Prasad--Sommerfield (BPS) subsector of any $\mathcal N= 2$ 4D theory \cite{Gadde:2011uv} by a (non-unitary) vertex operator algebra (VOA) \cite{Beem:2013sza}. VOAs for $\mathcal N=3$ theories were initially constructed in \cite{Nishinaka:2016hbw,Lemos:2016xke} culminating in the work of \cite{Bonetti:2018fqz}. In that reference, it was conjectured that certain VOAs labelled by non-Coxeter crystallographic complex reflection groups encode the Schur subsector of the known $\mathcal N=3$ S-fold theories. In particular, \cite{Bonetti:2018fqz} gave a prescription for an elegant free-field realisation of such VOAs, along the lines of \cite{Adamovic:2014lra}. By constructing the latter, one is able to recover the ``Macdonald'' limit of the superconformal index \cite{Gadde:2011uv} for $\mathcal N=3$ S-fold theories, from the VOA vacuum character. See also \cite{Song:2016yfd} for an alternate prescription on implementing a Macdonald grading of the chiral algebra.

Albeit concrete, implementing the findings of \cite{Bonetti:2018fqz} in practice quickly becomes computationally intensive. It is difficult to write down the explicit free-field realisation of the relevant VOAs in all but the simplest of cases, and also to evaluate the corresponding vacuum characters in a fugacity expansion for increasing conformal weights. The goal of this short note is to show how far one can get by implementing a brute-force approach using mathematical software, for the VOAs labelled by the complex reflection groups $G(k,1,1) = \mathbb Z_k$, $k = 3,4,6$ ($\mathbb Z_{3,4}$ label rank-one S-fold models) and $G(3,1,2)$ (labels a rank-two S-fold model). We employ the $G(k,p,N)$ notation of \cite{shephard_todd_1954}, where $p$ is a divisor of $k$; general complex reflection groups are denoted as $\mathsf G$.

Towards that end, we reconstruct and explicitly exhibit the free-field realisations of \cite{Bonetti:2018fqz} for the theories of interest. We then provide algorithms for automating the process of finding null states and for evaluating the VOA vacuum characters. Our code, appended to this letter, can in principle be executed to obtain the corresponding Macdonald index at arbitrary orders in a fugacity expansion. Note however that the vacuum character computation time increases exponentially as a function of the conformal weight. Our code is also customisable---and we have clearly signposted how to do so---for the reader interested in extending it to the evaluation of vacuum characters for VOAs labelled by other complex reflection groups, once the complete free-field realisation of the VOA has been found. 

Our results, all of which have been collected in the ancillary ``vacuum\_characters\_summary.nb'' for quick reference, can be used to check the conjecture of \cite{Bonetti:2018fqz} against independent calculations of the Macdonald index of 4D $\mathcal N=3$ S-fold theories and vice versa. For example, a proposal for the Schur limit of the superconformal index---a special case of the Macdonald index---for the rank-one $\mathbb Z_3$ S-fold theory was put forward in \cite{Zafrir:2020epd}. In that limit, their and our findings are in complete agreement.

\section{From Vacuum Characters to Indices}

We begin with a minimal introduction to the results of \cite{Bonetti:2018fqz}, to which we refer the reader for a full account. 4D $\mathcal N=2$ SCFTs, $\mathcal T$, contain a BPS subsector that is isomorphic to a VOA, $\chi[\mathcal T]$ \cite{Beem:2014zpa}. In this correspondence, the VOA central charge $c$ is related to the type-B Weyl-anomaly coefficient in four dimensions as $c = -12 c_{\mathrm{4D}}$, while the VOA vacuum character reproduces the Schur limit of the 4D superconformal index. The Schur subsector of pure $\mathcal N=3$ 4D S-fold SCFTs was conjectured to be isomorphic to ``$\mathcal N=2$ VOAs''---i.e. VOAs containing the 2D $\mathcal N=2$ superconformal algebra (SCA) as a subalgebra---$\mathcal W_{\mathsf G}$ labelled by non-Coxeter crystallographic complex reflection groups. In fact, \cite{Bonetti:2018fqz} proposed a free-field realisation of $\mathcal W_{\mathsf G}$ in terms of a subalgebra of $ \textrm{rank}(\mathsf G)$  copies of the $\beta\gamma b c $ ghost system. This subalgebra was identified with the kernel of a screening operator, $\mathbb S = \int \mathsf J $, acting on the $\beta\gamma bc$ systems. The action of $\mathbb S$ is defined as $\mathbb S \cdot X = \{\mathsf J X\}_1$, where $\{\mathsf J X\}_1$ denotes the coefficient of the order-one pole in the holomorphic OPE of the screening current $\mathsf J$ with some operator $X$.

A nice feature of the free-field description, which we will use extensively in the evaluation of the vacuum characters, is that null states built out of strong generators are identically zero. The free-field realisation also facilitates the introduction of a corresponding R-filtration, inherited from the R-symmetry of the 4D $\mathcal N=2$ SCFT.

The vacuum character of the R-filtered VOA is 
\begin{align}\label{vacch}
  \chi_{{\mathcal W}_\mathcal{G}}(q,\xi,z) := \mathrm{Tr}  (-1)^{F} q^{h} \xi^{R + r}z^m\;,
\end{align}
where $F$ is the fermion number, $h$ is the conformal dimension, and $r,m$ are associated with the $\mathsf{gl(1)}$ outer automorphism and $\mathsf{gl(1)}$ subalgebra of the 2D $\mathcal N=2 $ SCA $\mathsf{osp(2|2)}$ respectively. $R$ is the weight under the R-filtration and the vacuum character is normalised so as to start with a ``1'' in its $q$ expansion. This vacuum character can be further refined by taking $\xi^{R + r} \to y^R v^r$. In the free-field realisation the $\beta\gamma b c$ fields carry the quantum numbers presented in Tab.~\ref{tab:1}.
\begin{table}
  \renewcommand{\arraystretch}{1.3}
  \begin{tabular}{|c|c|c|c|c|}
\hline & $h$ & $m$  & $r$ & $R$\\
    \hline \hline $\beta_{\ell}$ & $\frac{1}{2} p_{\ell}$ & $\frac{1}{2} p_{\ell}$ & 0 & $\frac{1}{2} p_\ell$\\
\hline $b_{\ell}$ & $\frac{1}{2}\left(p_{\ell}+1\right)$ & $\frac{1}{2}\left(p_{\ell}-1\right)$ & +$\frac{1}{2}$& $1- \frac{1}{2}p_\ell$ \\
\hline $c_{\ell}$ & $-\frac{1}{2}\left(p_{\ell}-1\right)$ & $-\frac{1}{2}\left(p_{\ell}-1\right)$ &  $-\frac{1}{2}$ & $\frac{1}{2} p_\ell$\\
\hline $\gamma_{\ell}$ & $1-\frac{1}{2} p_{\ell}$ & $-\frac{1}{2} p_{\ell}$ & 0 & $1-\frac{1}{2} p_\ell$\\
\hline $\partial$ & 1 & 0  & 0 & 0 \\
\hline
  \end{tabular}
  \caption{\label{tab:1} Quantum numbers for the $\beta_\ell\gamma_\ell b_\ell c_\ell $ ghost systems used in VOA free-field realisations. The $\ell = 1,\ldots,\mathrm{rank}(\mathsf G)$ labels the ghost-system species and $p_\ell$ are the degrees of the fundamental invariants of $\mathsf G$. These are given for $\mathbb Z_k$ by $p_1= k$ and for $G(3,1,2)$ by $(p_1, p_2) = (3,6)$.}
\end{table}

Eq.~\eqref{vacch} was conjectured to correspond to the Macdonald limit of the 4D superconformal index of a theory for which $\mathcal W_{\mathsf{G}}$ is the associated VOA, $\mathcal W_{\mathsf{G}} =  \chi[\mathcal T ]$. This is defined through \cite{Gadde:2011uv}
\begin{align}
  \mathcal I^{\mathcal T}_{\mathrm{Macdonald}}(q,t,z) := \mathrm{Tr} (-1)^F q^{E - 2R - r} t^{R + r}z^m\;,
\end{align}
where the trace is taken over the set of Schur operators of the 4D SCFT. Here $E$ is the 4D conformal dimension, while $R$ and $r$ are charges for the Cartan generators of the $SU(2)_R$ and $U(1)_r$ R-symmetry groups respectively. Note that while from the point of view of this 4D $\mathcal N=2$ Macdonald index $m$ is a quantum number for a global $U(1)_F$, in the full $\mathcal N=3$ description it is part of the $U(3)\supset SU(2)_R \times U(1)_r \times U(1)_F$ R-symmetry group. We should emphasise that in order to connect with \eqref{vacch} one needs to redefine $t \to \xi q$ so that
\begin{align}\label{mcd}
  \mathcal I^{\mathcal T}_{\mathrm{Macdonald}}(q,\xi,z) = \mathrm{Tr} (-1)^F q^{E - R} \xi^{R + r}z^m\;.
\end{align}

\section{Implementation}

We now describe the strategy behind our code, while detailed results for each case are presented in subsequent subsections. We are interested in the evaluation of the vacuum character for VOAs labelled by crystallographic, non-Coxeter complex reflection groups $\mathbb Z_{3,4,6}$ and $G(3,1,2)$, and interpreting them as Macdonald indices for 4D $\mathcal N=3$ S-fold theories. To do so one needs to consider \eqref{vacch} and trace over all the states created by acting only with normal-ordered products and derivatives of the strong generators of the VOA on the $\mathsf{sl(2)}$-invariant vacuum, up to a given conformal weight, while removing the contributions from null states. To speed up the calculation, we have taken advantage of the symmetry of the spectrum under conjugation, by only constructing positively-charged states under the two $\mathsf{gl(1)}$ symmetries of the VOA.

The identification of null states is usually laborious and this is where the free-field realisation becomes helpful. The construction algorithm of \cite{Bonetti:2018fqz} starts with a simple free-field prescription for the generators of the 2D $\mathcal N=2$ SCA and the chiral strong generators of the $\mathcal W_{\mathsf G}$-algebra. One then introduces an ansatz for the remaining bosonic strong generators constructed out of all possible super-Virasoro primary, free-field combinations with the requisite quantum numbers and undetermined coefficients, and fixes the latter by closing the VOA under the OPE.

We have diverged slightly from this recipe in the following way. For the VOAs $\chi[\mathcal T]$ that have appeared thus far in the literature, the OPE coefficients can be completely fixed by writing down the most general expressions for the expected set of generators and then imposing associativity by solving the Jacobi identities \cite{Lemos:2016xke}. The W-algebra presentations of the $\mathcal W_{\mathsf G}$ VOAs of interest to us were already given in \cite{Bonetti:2018fqz} and inserting the free-field ansatze into the existing OPEs straightforwardly fixes the undetermined coefficients for all remaining strong generators (using the OPEdefs \cite{Thielemans:1994er} and/or SOPEN2defs \cite{Krivonos:1995bk} Mathematica packages); we will expand upon this point on a case-by-case basis when discussing our results.

Equipped with this information we proceed to our main algorithm. In summary, all null states at a given conformal weight can be identified by looking for all possible combinations of states with the same quantum numbers that are identically zero upon using the free-field realisation. This last step requires manipulating normal-ordered products of $\beta \gamma bc$ ghosts, making heavy use of the OPEdefs  \cite{Thielemans:1994er} and ope.math \cite{Fujitsu:1994np} Mathematica packages. The null states that we find contain all those predicted in \cite{Bonetti:2018fqz}. It is then straightforward to write down the vacuum character for given values of quantum numbers. Note that by unrefining in the $R$ fugacity (e.g. if one were interested in the Schur index) the algorithm becomes faster; we have a dedicated section in our code for this special case. 

\begin{figure}
\includegraphics[scale=.5]{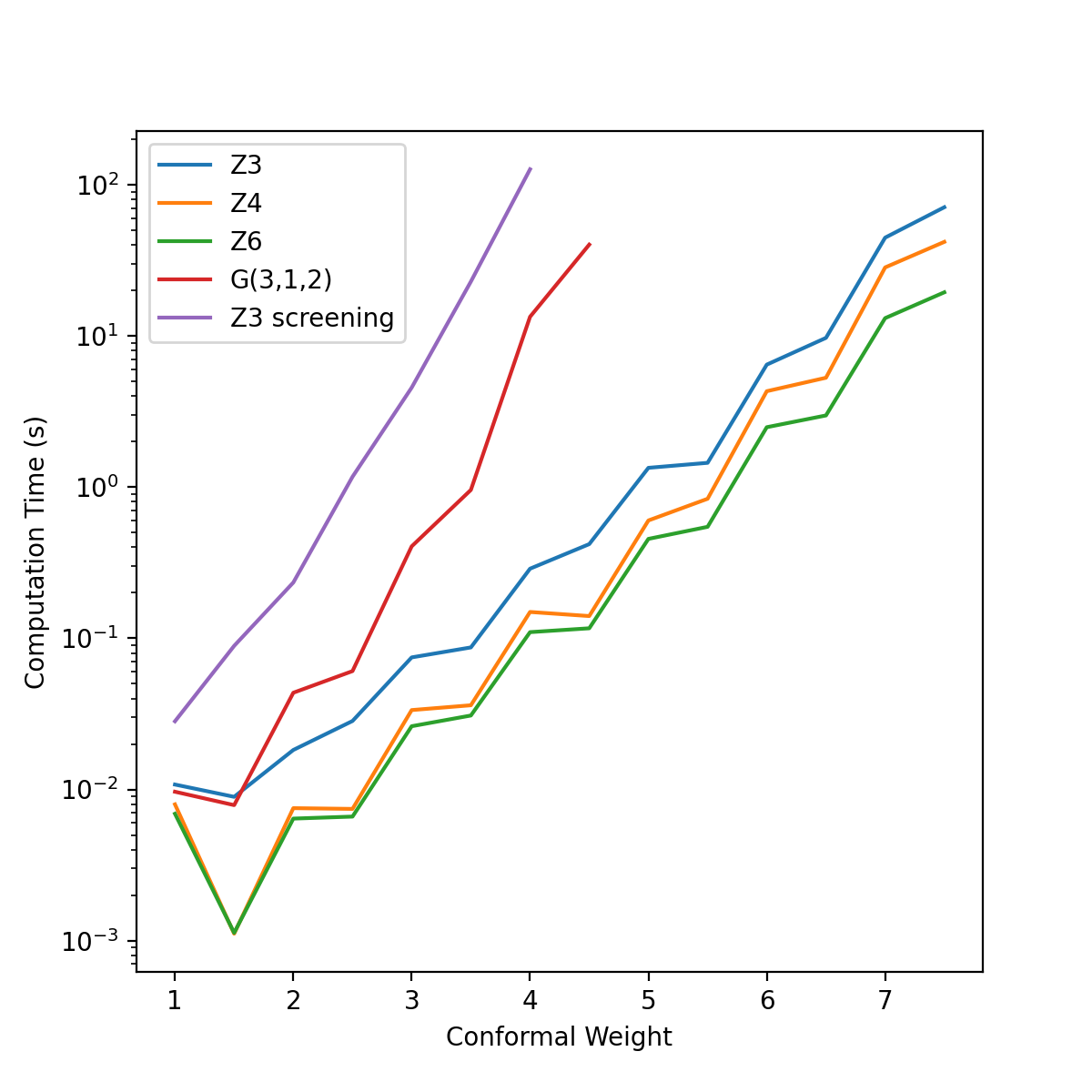}
  \caption{\label{fig:1} Computation times for carrying out the calculation of the vacuum character for VOAs at different conformal weights. Several different VOAs are shown on the graph for comparison, along with the $\mathbb Z_3$ theory computed using the kernel of the screening operator. We used a desktop PC with an Intel Core i7-6700K CPU clocked at 4GHz, and 32GB RAM.}
 \end{figure}

In addition to the above method, we have cross-checked our refined $\mathbb Z_{3,4}$ results using a second algorithm that makes no connection to the W-algebra presentation. This procedure constructs the VOA spectrum using all states in $\mathrm{rank}(\mathsf G)$ copies of the $\beta\gamma bc $ ghost system that lie in the kernel of a screening operator, $\mathbb S$. For $\mathsf G = \mathbb Z_{k}$ this is given by \cite{Bonetti:2018fqz}
\begin{align}
  \mathsf{J}=b \; e^{\left(k^{-1}-1\right)(\chi+\phi)}\;,
\end{align}
where $\chi, \phi$ are chiral bosons
\begin{align}
  \beta=e^{\chi+\phi}, \quad \gamma=\partial \chi\; e^{-\chi-\phi}\;,
\end{align}
and all expressions should be considered as normal-ordered. Such an approach is conceptually more straightforward---construct all states using free fields and then keep those in the kernel of $\mathbb S$---but is computationally more expensive as can be seen from Fig~\ref{fig:1}. E.g. at $h = 9/2$ one already needs to check 941 and 881 terms for $\mathbb Z_3$ and $\mathbb Z_4$ respectively. 

\section{Results: $\mathsf G = \mathbb Z_3$}

This is a rank-one VOA with central charge $c = -15$. Its W-algebra presentation involves the following strong generators: $\mathcal T$, $\mathcal J$, $\mathcal G$ and $\widetilde{\mathcal G}$ from the 2D $\mathcal N=2$ SCA, as well as the chiral and anti-chiral generators $\mathcal W_3$, $\overbar{\mathcal  W}_3$ and their superpartners $\mathcal G_{\mathcal W_3}$ and $\widetilde{\mathcal G}_{\overbar{\mathcal  W}_3}$. Here $\mathcal G_{\mathcal W_3}:= \{\mathcal G \mathcal W_3\}_1$ and so on. For the explicit free-field realisation in terms of a single $\beta\gamma b c$ ghost system one starts with a prescription for the $\mathcal N=2$ SCA generators as well as for $\mathcal W_3$, $\mathcal G_{\mathcal W_3}$. The ansatz for $\overbar{\mathcal  W}_3$ contains 8 undetermined coefficients. We always count these before imposing the super-Virasoro primary constraint. Through the OPEs from the W-algebra presentation one can use it to also determine the free-field realisation of $\widetilde{\mathcal G}_{\overbar{\mathcal  W}_3}$. We have calculated the fully-refined vacuum character \eqref{vacch} up to $O(q^{8})$ with the accompanying ``null states.nb'' and have cross-checked this result using the screening-operator approach in ``screening.nb'' up to $O(q^{4})$. The full expression can be found in ``vacuum\_characters\_summary.nb''. 

This VOA is expected to encode the Schur sector of a rank one, 4D S-fold $\mathcal N=3$ SCFT, with a Coulomb-branch operator of dimension $\Delta = 3$ and trace-anomaly coefficient $c_{\mathrm{4D}} = \frac{5}{4}$. Through \eqref{mcd} the Macdonald index of this S-fold theory---including the global $U(1)_F$ fugacity---can be identified with the refined vacuum character. Below we only present the simpler, Schur limit of these expressions for brevity, where $\xi\to 1$, $z\to 1$. Then:
\begin{align}\label{schurz3}
  \mathcal I^{\mathbb Z_3}_{\mathrm{Schur}} &=  1+q+q^2+2 q^3-2 q^{7/2}+3 q^4-2 q^{9/2}+4 q^5-4 q^{11/2}\nonumber \\
                                              & +6 q^6-6 q^{13/2}+8 q^7-8 q^{15/2}+11 q^8 + O(q^{17/2}).
\end{align}
It is interesting to observe that the Schur index for this theory matches the expansion of the following closed-form expression up to $O(q^{10})$, although we currently have neither a derivation for it nor a justification for why it should hold to all orders:
\begin{align}
  \frac{1}{3}\sum_{\epsilon\in \mathbb Z_3}  \frac{\epsilon}{\sqrt{q}}&\text{P.E.}\left[\frac{1}{2}i_{\mathcal N=4}(q)\left(\epsilon+\frac{1}{\epsilon}\right)\right]\;.
\label{ho}
\end{align}
Here $\text{P.E.}[f(z)]:=\exp[\sum_{n=1}^{+\infty}\frac{1}{n}f(z^n)]$  is the plethystic exponential, while $i_{\mathcal N=4}(q) = \frac{2 q^{\frac{1}{2}} (1- q^{\frac{1}{2}})}{1-q}$ coincides with the single-letter Schur index of $\mathcal N=4$ super-Yang--Mills.

In \cite{Zafrir:2020epd}, an independent argument for determining the Schur index of the $\mathsf G = \mathbb Z_3$ S-fold theory was presented. This entailed starting from an $\mathcal N=1$ 4D UV Lagrangian theory, and flowing to an interacting  $\mathcal N=1$ SCFT in the IR, which can also be reached from the $\mathbb Z_3$ S-fold theory via an $\mathcal N=1$ preserving marginal deformation. Upon relabelling $q\to p^2$,  our result \eqref{schurz3} agrees with that of \cite{Zafrir:2020epd}---listed to $O(q^7)$---providing a strong consistency check of both calculations.

\section{Results: $\mathsf G = \mathbb Z_4$}

This is a rank-one VOA with central charge $c = -21$. Its W-algebra presentation involves the following strong generators: $\mathcal T$, $\mathcal J$, $\mathcal G$ and $\widetilde{\mathcal G}$ from the 2D $\mathcal N=2$ SCA, as well as the chiral and anti-chiral generators $\mathcal W_4$, $\overbar{\mathcal W}_4$ and their superpartners $\mathcal G_{\mathcal W_4}$ and $\widetilde{\mathcal G}_{\overbar{\mathcal W}_4}$. For the explicit free-field realisation in terms of a single $\beta\gamma b c$ ghost system one starts with a prescription for the $\mathcal N=2$ SCA generators as well as for $\mathcal W_4$, $\mathcal G_{\mathcal W_4}$. The ansatz for $\overbar{\mathcal W}_4$ contains 19 undetermined coefficients. Through the OPEs in the W-algebra presentation one can use it to also determine the free-field realisation of $\widetilde{\mathcal G}_{\overbar{\mathcal W}_4}$. We have calculated the fully-refined vacuum character \eqref{vacch} up to $O(q^{8})$ with the accompanying ``null states.nb'' and exhibited it in ``vacuum\_characters\_summary.nb''. We have also cross-checked this result using the screening-operator approach in  ``screening.nb'' up to $O(q^4)$.

This VOA is expected to encode the Schur sector of a rank one 4D S-fold $\mathcal N=3$ SCFT, with a Coulomb-branch operator of dimension $\Delta = 4$ and trace-anomaly coefficient $c_{\mathrm{4D}} = \frac{7}{4}$. Through \eqref{mcd} the Macdonald index of this S-fold theory can be identified with the vacuum character. The Schur limit of these expressions yields:
\begin{align}
  \mathcal I^{\mathbb Z_4}_{\mathrm{Schur}} &=  1+q-2 q^{3/2}+5 q^2-6 q^{5/2}+10 q^3-16 q^{7/2}+27 q^4\cr
  & -38 q^{9/2}+56 q^5-86 q^{11/2}+129 q^6-178 q^{13/2}\nonumber \\
    & +251 q^7-362 q^{15/2}+511 q^8+ O(q^{17/2})\;.
\end{align}
In this case the W-algebra construction is such that the bosonic states always appear with integer while the fermionic ones with half-integer conformal weights. Therefore there are no cancellations between bosonic and fermionic states at each level and the chiral algebra partition function reproduces the partition function of Schur operators in the corresponding 4D $\mathcal N=3$ theory.

\section{Results: $\mathsf G = \mathbb Z_6$}

This is a rank-one VOA with central charge $c = -33$. Its W-algebra presentation involves the following strong generators: $\mathcal T$, $\mathcal J$, $\mathcal G$ and $\widetilde{\mathcal G}$ from the $\mathcal N=2$ SCA, as well as the chiral and anti-chiral generators $\mathcal W_6$, $\overbar{\mathcal W}_6$ and their superpartners $\mathcal G_{\mathcal W_6}$ and $\widetilde{\mathcal G}_{\overbar{\mathcal W}_6}$. For the explicit free-field realisation in terms of a single $\beta\gamma b c$ ghost system one starts with a prescription for the 2D $\mathcal N=2$ SCA generators as well as for $\mathcal W_6$, $\mathcal G_{\mathcal W_6}$. The ansatz for $\overbar{\mathcal W}_6$ contains 87 undetermined coefficients. Through the OPEs in the W-algebra presentation one can use it to also determine the free-field realisation of $\widetilde{\mathcal G}_{\overbar{\mathcal W}_6}$. We have calculated the fully-refined vacuum character \eqref{vacch} up to $O(q^{8})$ with the accompanying ``nullstates.nb''.  The full result can be found in ``vacuum\_characters\_summary.nb''.  In the limit $\xi\to 1$, $z\to 1$ this reads:
\begin{align}
  \chi_{\mathcal W_{\mathbb Z_6}}&=  1+q-2 q^{3/2}+3 q^2-4 q^{5/2}+8 q^3-12 q^{7/2}+19 q^4\cr
  & -26 q^{9/2}+38 q^5-58 q^{11/2}+85 q^6 -116 q^{13/2}\nonumber \\
    & +165 q^7-236 q^{15/2}+326 q^8+ O(q^{17/2}) \;.
\end{align}
No known S-fold theory is associated with this VOA \cite{Aharony:2016kai}.

\section{Results: $\mathsf G =  G(3,1,2)$}

This is our only rank 2 example, with central charge $c = -48$. Its W-algebra presentation involves the following strong generators: $\mathcal T$, $\mathcal J$, $\mathcal G$ and $\widetilde{\mathcal G}$ from the 2D $\mathcal N=2$ SCA,  the chiral and anti-chiral generators $\mathcal  W_3$, $\mathcal W_6$, $\overbar{\mathcal W}_3$, $\overbar{\mathcal W}_6$, plus their superpartners $\mathcal G_{\mathcal W_3}$, $\mathcal G_{\mathcal W_6}$ and $\widetilde{\mathcal G}_{\overbar{\mathcal W}_3}$, $\widetilde{\mathcal G}_{\overbar{\mathcal W}_6}$, as well as $\mathcal U$---which is self conjugate---and its superpartners  $\mathcal G_{\mathcal U}$, $\widetilde{\mathcal G}_{\mathcal U}$ and $\mathcal G_{\widetilde{\mathcal G}_{\mathcal U}}$. One also needs non-chiral $\mathcal O$, $\overbar{\mathcal O}$ and their superpartners $\mathcal G_{\mathcal O }$, $\widetilde{\mathcal G}_{\mathcal O }$, $\mathcal G_{\overbar{\mathcal O} }$, $\widetilde{\mathcal G}_{\overbar{\mathcal O }}$, $\mathcal G_{\widetilde{\mathcal G}_{\mathcal O }}$, $\mathcal G_{\widetilde{\mathcal G}_{\overbar{\mathcal O} }}$.

The free-field realisation requires two ghost systems, $\beta_\ell\gamma_\ell b_\ell c_\ell$ with $\ell = 1,2$. One starts with a prescription for the 2D $\mathcal N=2$ SCA generators as well as for $\mathcal W_3$, $\mathcal W_6$, $\mathcal G_{\mathcal W_3}$ and $\mathcal G_{\mathcal W_6}$. The ansatz for $\overbar{\mathcal W}_3$ contains 84 undetermined coefficients. It turns out that through the OPEs in the W-algebra presentation,  one can fix the coefficients of $\overbar{\mathcal W}_3$ and by doing so also determine the free-field realisation of all remaining generators. We have calculated the fully-refined vacuum character \eqref{vacch} up to $O(q^{9/2})$ with the accompanying ``null states.nb''.  The full result can be found in ``vacuum\_characters\_summary.nb''.

This VOA is expected to encode the Schur sector of a rank-two 4D S-fold $\mathcal N=3$ SCFT, with Coulomb-branch operators of dimension $\Delta = 3,6$ and trace anomaly coefficient $c_{\mathrm{4D}} = 4$. Through \eqref{mcd} the Macdonald index of this S-fold theory can be identified with the vacuum character. If for simplicity one considers  the limit $z\to 1$:
\begin{align}
  \mathcal I^{G(3,1,2)}_{\mathrm{Macdonald}} & =  1+q \xi +q^{3/2} \left(-\sqrt{\xi }+\xi ^{3/2}\right)+q^2 \left(\xi +\xi ^2\right)\cr
  & +q^{5/2} \left(-\sqrt{\xi }-\xi ^{3/2}+2 \xi ^{5/2}\right)+q^3 \left(\xi -\xi ^2+2
    \xi ^3\right)\cr
  & +q^{7/2} \left(-\sqrt{\xi }-2 \xi ^{3/2}+2 \xi ^{5/2}+\xi ^{7/2}\right)\cr
  & +q^4 \left(2 \xi -\xi ^3+3 \xi ^4\right)\cr
    & +q^{9/2} \left(-\sqrt{\xi }-3 \xi
      ^{3/2}+\xi ^{5/2}+\xi ^{7/2}+2 \xi ^{9/2}\right) \cr
      & + O(q^5)\;.
\end{align}
\newpage

\section{Conclusions}

In this letter we have calculated vacuum characters of rank-one and rank-two VOAs labelled by non-Coxeter, crystallographic complex reflection groups. This involved a brute-force implementation of the algorithms presented in \cite{Bonetti:2018fqz} and leads to the Macdonald index of certain 4D $\mathcal N=3$ S-fold SCFTs. Our results were given as an expansion in the fugacity that keeps track of the conformal weight, and were truncated to orders that require short computation times when using a desktop computer; they can be pushed to arbitrary higher orders by allocating appropriate resources. As they stand, they can already be used as new data for $\mathcal N=3$ SCFTs. E.g. the $\mathsf G = \mathbb Z_3$ result agrees in the Schur limit with \cite{Zafrir:2020epd}.

Our code is customisable. We have clearly signposted where changes would need to be made to return vacuum characters of VOAs labelled by different complex reflection groups, for which the free-field realisation is known. In particular, it would be very interesting to extend this approach to the  $\mathcal N=3$ S-fold SCFT of rank two associated with $G(4,1,2)$ and the rank-three example $G(3,3,3)$; a proposal for the Schur index of the latter was also given in \cite{Zafrir:2020epd}. Unfortunately, finding the free-field realisation for both these VOAs---already needed before identifying the null states---is a challenging task: the simplest anti-chiral strong generator ansatze involve 425 and 2265 undetermined coefficients respectively. It would perhaps be more promising to use the screening-operator approach, upon determining $\mathbb S$. Although our screening-operator code is currently more expensive to run, it could benefit from optimisations that parallelise the computations, hence making it significantly faster on multi-core clusters. It will also be interesting to check these results by directly studying the BPS-states of $\mathcal N=3$ theories. One way to do so would be to study three-string junctions in S-fold backgrounds as  in \cite{Agarwal:2016rvx}. We hope to return to some of these questions in the near future.


\section{Acknowledgements}

We would like to thank B.~Ergun, T.~Fran\c ca, M.~Martone, P.~Longhi, C.~Meneghelli, A.~Pini, E.~Pomoni, F.~Schaposnik
Massolo and J.~Song for useful discussions and comments. We acknowledge financial support by the Royal Society URF\textbackslash R\textbackslash 180009, RGF\textbackslash EA\textbackslash 181049, RGF\textbackslash EA\textbackslash 180073 and the STFC ST/P000754/1.

\bibliography{4DN3}
\bibliographystyle{apsrev}

\end{document}